\documentclass[12pt,preprint]{aastex}

\slugcomment{Accepted for publication by The Astrophysical Journal}
\shorttitle{Role of coherent structures}
\shortauthors{Arzner et al.}

\begin{document}

\def\mlambda{\mbox{\boldmath $\lambda$}}                        % boldface Greek letters
\def\mxi{\mbox{\boldmath $\xi$}}
\def\mphi{\mbox{\boldmath $\phi$}}
\def\mkappa{\mbox{\boldmath $\kappa$}}
\def\meta{\mbox{\boldmath $\eta$}}
\def\mhdeb{({\bf {\cal E}}{\bf {\cal B}})}
\def\mhdebd{({\bf {\cal E}}{\bf {\cal B}})_{DNS}}
\def\mhdebs{({\bf {\cal E}}{\bf {\cal B}})_{STO}}
\def\bfe{{\bf e}}
\def\bff{{\bf f}}
\def\bfb{{\bf b}}
\def\bfj{{\bf j}}
\def\bfk{{\bf k}}
\def\bfr{{\bf r}}
\def\bfu{{\bf u}}
\def\bfv{{\bf v}}
\def\bfp{{\bf p}}
\def\bfx{{\bf x}}
\def\bfy{{\bf y}}
\def\bfz{{\bf z}}
\def\bhat{{\bf \hat{b}}}
\def\ehat{{\bf \hat{e}}}
\def\jhat{{\bf \hat{j}}}
\def\uhat{{\bf \hat{u}}}
\def\vhat{{\bf \hat{v}}}
\def\llangle{\langle\hspace{-2pt}\langle}
\def\rrangle{\rangle\hspace{-2pt}\rangle}

\title{The Effect of Coherent Structures on Stochastic Acceleration in MHD Turbulence}
\author{Kaspar Arzner}
\affil{Paul Scherrer Institut, Villigen, CH-5232 Villigen PSI, Switzerland}
\email{arzner@astro.phys.ethz.ch}

\author{Bernard Knaepen, Daniele Carati, and Nicolas Denewet}
\affil{Universit\'e libre de Bruxelles, Bruxelles, Belgium}
\email{bknaepen@ulb.ac.be}

\author{Loukas Vlahos}
\affil{Department of Physics, Aristotle University of Thessaloniki,
        54124 Thessaloniki, Greece}
\email{vlahos@astro.auth.gr}

\begin{abstract}
We investigate the influence of coherent structures on particle acceleration
in the strongly turbulent solar corona. By randomizing the Fourier phases of a pseudo-spectral
simulation of isotropic MHD turbulence (Re $\sim 300$), and tracing 
collisionless test protons in both the exact-MHD and phase-randomized fields, it is found that 
the phase correlations enhance the acceleration efficiency during the first
adiabatic stage of the acceleration process. The underlying physical mechanism is identified as the 
dynamical MHD alignment of the magnetic field with the electric current, which favours parallel 
(resistive) electric fields responsible for initial injection. 
Conversely, the alignment of the magnetic field with the bulk velocity
weakens the acceleration by convective electric fields $- \bfu \times \bfb$ at a non-adiabatic 
stage of the acceleration process.
We point out that non-physical parallel electric fields in random-phase turbulence proxies
lead to artificial acceleration, and that the dynamical MHD alignment can be taken into account 
on the level of the joint two-point function of the magnetic and electric fields, and is 
therefore amenable to Fokker-Planck descriptions of stochastic acceleration.
\end{abstract}

\keywords{acceleration of particles --- MHD --- methods: numerical}

\section{INTRODUCTION}

Turbulent electromagnetic fields can act as particle accelerators in astrophysical situations such 
as solar flares \citep{miller97,benz03}. The turbulence is governed by non-linear fluid equations, and the
turbulent fields therefore deviate from a simple Gaussian field \citep{adler81}, where the Fourier phases are 
at random, and the field is completely specified by its two-point function. In real space, the turbulent phase
correlations manifest in coherent structures, with alignment of magnetic and velocity fields \citep{pouquet86},
and in non-Gaussian higher-order correlations \citep{mccomb90,falkovich01}.

The latter are not captured by traditional stochastic acceleration theory, which invokes the two-point functions 
of the magnetic field to predict the particle diffusion coefficients (e.g., \cite{urch77,achterberg81,karimabadi90,
urch91,bykov93,schlickeiser98,schlickeiser02}).
On the numerical side, some studies (e.g., \citealt{mace00, paesold03, arzner04}) have worked with random-phase
turbulence proxies and therefore neglected coherent structures, while others (e.g., \citealt{matthaeus84, dmitruk03})
used simulated MHD fields and took them into account.

The presence of coherent structures may have conflicting effects on the acceleration efficiency \citep{dmitruk03}. 
On the one hand, the coherent structures may act as traps and detain particles on their way to high energies.
On the other hand, coherent structures may host large electric fields which help acceleration.
Which of the two effects prevails is hard to predict on theoretical grounds and depends on the topology and 
size of the coherent structures. A numerical exploration is thus needed to capture the whole diversity of 
possible orbit behaviour. The present article reports on relativistic test particle simulations in turbulent MHD 
electromagnetic fields, following a similar approach as \cite{dmitruk03}. By applying an artificial phase randomization
to the electromagnetic fields we then study the effect of coherent structures on test particle acceleration, and
compare it with the Gaussian field limit. 

The astrophysical realization envisaged by the present simulation is proton acceleration in solar flares, with 
emphasis on the high-energy regime where adiabaticity of the orbit breaks down. Scaling to high-energy 
electrons or heavier ions is in principle possible, but not explicitly discussed. The paper is organized as follows. 
Section \ref{field_sect} describes the MHD fields and their randomization, and Section \ref{orbit_sect} describes 
the numerical generation of particle orbits and the physical approximations made in doing so. Section \ref{result_sect}
presents the particle results, followed by a discussion of the acceleration mechanism (Sect. \ref{discussion_sect}).
Section \ref{summary_sect} contains a summary of the simulation results, and a discussion of 
their validity and implications.

\section{\label{field_sect}ELECTROMAGNETIC FIELDS}

\subsection{Random Versus Coherent Fields}

As indicated in the introduction, the focus of this paper is on the role
of turbulent coherent structures on particle acceleration. Here we assume that particles are only subject to an 
electric field $\bfe$ and a magnetic field $\bfb$ through the (relativistic) Lorentz force
(we will refer to a couple $\bfe $ and $\bfb$ as an ``accelerator''). The electric field is taken according
to MHD, 
%
%\begin{align}
%\begin{equation}
\begin{eqnarray}
\bfe = - \bfu \times \bfb + \eta {\bf j} \, , \;\;\; {\bf \nabla} \times {\bf b}  = {\bf j} \, ,   \label{je}
\end{eqnarray}
%\end{equation}
%\end{align}
%
where $\bfu$ is the bulk velocity, $\bfj$ is the electric current and $\eta$ is the magnetic diffusivity. In Eq. 
(\ref{je}) and from now on we use Alfv\'en units, i.e., the magnetic field is scaled to the Alfv\'en velocity,
$\bfb = {\bf B}/\sqrt{\mu_0 \rho}$, and measured in the same units as the bulk velocity.
% see e.g. Ohsaki and Mahajan, Phys. of Plasmas 11, 898 (2003)
Another assumption of the present study is that the
particles' motion take place in a cubic domain with periodic boundary conditions.
This implies that at any instant the MHD fields can be viewed as 
superpositions of waves and decomposed
in Fourier modes:
%
%\begin{align}
%\begin{equation}
\begin{eqnarray}
\bfu(\bfx)&=&\sum  \tilde \bfu({\bf k}) e^{i{\bf k}\cdot \bfx},\\
\bfb(\bfx)&=&\sum  \tilde \bfb({\bf k}) e^{i{\bf k}\cdot \bfx}.
\end{eqnarray}
%\end{equation}
%\end{align}
%
In the above expansions, $\tilde \bfu({\bf k})$ and $\tilde \bfb({\bf k})$ are 
complex numbers which satisfy, $\tilde \bfu(-{\bf k})=\tilde \bfu^*({\bf k})$, 
$\tilde \bfb(-{\bf k})=\tilde \bfb^*({\bf k})$ and can be written as,
%
%\begin{align}
%\begin{equation}
\begin{eqnarray}
\tilde \bfu({\bf k})= \tilde \bfu_r({\bf k}) e^{i{\alpha}({\bf k})},\quad
\tilde \bfb({\bf k})= \tilde \bfb_r({\bf k}) e^{i{\beta}({\bf k})} \label{ubf}
\end{eqnarray}
%\end{equation}
%\end{align}
%
with real-valued $\bfu_r$, $\bfb_r$, $\alpha$ and $\beta$.
Constructing a numerical field like (\ref{ubf}) thus requires
information both on the modulus and the phases of its Fourier modes. In
homogeneous and isotropic turbulence, the modulus of Fourier modes are
usually set to given values that depend only on the norm of the
corresponding wave vector to mimic a given energy spectra. Assigning
appropriate phases to the Fourier modes is a much more complex task. In
a real turbulent field, these phases have to be correlated in such a way
as to produce a field with subtle and intricate structures. Prescribing
them numerically represents the same complexity as prescribing the
electromagnetic field at each point in space and is thus not practical.
The most simple way of circumventing this problem is to assign random
numbers in the interval $\left[-\pi, \pi\right]$ to the phases and impose
a given distribution, most often the uniform distribution to respect isotropy.
The fields generated can then be qualified as stochastic but still have
a prescribed energy spectra. More realistic phases can be obtained by
generating the electromagnetic field through direct numerical
simulations (DNS) of the MHD equations,
%
%\begin{align}
%\begin{equation}
\begin{eqnarray}
\partial_t u_i&=&-\partial_i (p/\rho) -u_j\partial_j
u_i+ b_j\partial_j b_i + \nu \Delta
u_i,\label{NSu}\\
\partial_t b_i &=&-u_j \partial_j b_i+ b_j
\partial_j u_i + \eta \Delta b_i,\label{NSb}
\end{eqnarray}
%\end{equation}
%\end{align}
%
where $p$ is the sum of the kinematic and magnetic pressures and $\nu$ is the 
kinematic viscosity. The magnetic field contains no mean contribution. Since we assume
incompressibility, $p$ is obtained by solving a Poisson equation and is
not an independent variable (e.g., \citealt{mccomb90}). Although
the initial condition used in (\ref{NSu}) and (\ref{NSb}) consists of
fields with random phases, it is observed that after some time,
the resulting electromagnetic field exhibits coherency that deviates
strongly from its random phases counterpart (e.g., \citealt{biskamp99}). Although the DNS approach
is very appealing, it should be stressed that it is very costly and that the values
of Reynolds numbers observed in typical turbulent systems are out of reach
of present-day supercomputers. In that sense, the fields obtained through DNS
also constitute models for real-life turbulence.

Our objective is here to study two kinds of accelerators: a random phases one 
and a coherent one obtained through DNS. To make things consistent, we construct
both accelerators in such a way that they have exactly the same spectra and
only differ through the value of the phases of their Fourier modes. As a shorthand,
the random phases accelerator will be denoted $\mhdebs$ while the DNS accelerator
will be denoted $\mhdebd$. In the next section we detail the construction of 
$\mhdebs$ and $\mhdebd$.

\subsection{MHD Simulations}

In order to obtain a suitable DNS electromagnetic field, we solve equations
(\ref{NSu}) and (\ref{NSb}) using a pseudo-spectral code in a cubic domain of
side length $l_x = l_y = l_z = 2\pi$ and impose periodic boundary conditions.
Our simulation code is dealiased and uses a 4th-order accurate Runge-Kutta 
time advancement scheme.
As explained above, the initial condition for $u_i$ and $b_i$ are random phases
fields of the form (\ref{ubf}) whose spectra match the one observed in the Comte-Bellot-Corrsin
grid turbulence experiment at stage 2 \citep{coco71} (this choice of spectra is motivated by
the fact that the corresponding turbulence can be resolved with a resolution of
$256^3$ Fourier modes). 
%The turbulence is then freely decaying in time \citep{biskamp99}.
The only extra requirement imposed on the Fourier modes is incompressibility, 
${\bf k} \cdot \tilde \bfb({\bf k}) = {\bf k} \cdot \tilde \bfu({\bf k})$ = 0. As time advances, 
the flow becomes ``physical" and we stop the simulation when the skewness of the velocity derivative 
$\partial u_1 / \partial x_1$ reaches a quasi constant value \citep{batchelor82,mccomb90,gotoh02,verma04}. 
This criteria is standard and indicates that turbulence has had time to develop. Here, the initialisation 
phase has taken about $2.5t_A$ where $t_A$ is the Alfv\'en time defined by
$t_A=L/v_A$, with $L$ the (magnetic or velocity) integral scale-length and 
$v_A^2 = \langle |\bfb|^2 \rangle /3$ (from here on, angular brackets $\langle ... \rangle$ 
denote a spatial average). Note that $v_A$ is determined by the (rms) magnetic fluctuations
and not by a background magnetic field. Some of the characteristics
of the flow at 2.5$t_A$ are listed in Table \ref{tableparamflow}. There is
approximate equipartition between $|\bfb|^2$ and $|\bfu|^2$, and both dissipate at
the same rate (magnetic Prandtl number $\sim$ 1), so that the magnetic and velocity characteristic 
scales are similar. The Reynolds number and magnetic Reynolds number are about 300.
The magnetic field extracted from the DNS at this final time and the electric field 
obtained through (\ref{je}) constitute the accelerator $\mhdebd$.
\begin{table}[h!]
\begin{center}
\begin{tabular}{l l l}
Resolution   &  \hspace{.7 cm} & $256^3$    \\
Box size ($l_x \times l_y \times l_z$) & & $2\pi \times 2\pi \times 2\pi$ \\ 
Rms velocity $u=\sqrt{\frac{\langle u_i u_i \rangle}{3}}$   &   &  2.20 \\
Rms magnetic field $b=\sqrt{\frac{\langle b_i b_i \rangle}{3}}$   &   &  2.39 \\
Viscosity $\nu$ &  &  0.006 \\
Diffusivity $\eta$ &  &  0.006 \\
Kinetic dissipation $\epsilon_u$ &  &  15.28\\
Magnetic dissipation $\epsilon_b$ &  &  20.57\\
Integral length-scale of u ($L_u = 3\pi/4 \times (\int \kappa^{-1} E_u(\kappa)d\kappa/
\int E_u(\kappa)d\kappa)$) &  &  0.79 \\
Integral length-scale of b ($L_b = 3\pi/4 \times (\int \kappa^{-1} E_b(\kappa)d\kappa/
\int E_b(\kappa)d\kappa)$) &  &  0.71 \\
%$Re=uL_u/\nu$ & & 290 \\
%$Re_b=uL_b/\eta$ & & 283 \\
Longitudinal Taylor microscale of u, $\lambda_u^2=15\nu u^2/\epsilon_u$ & & 0.169 \\
Longitudinal Taylor microscale of b, $\lambda_b^2=15\eta b^2/\epsilon_b$ & & 0.158\\
Eddy turnover time ($E_u/\epsilon_u$) & & 0.421
\end{tabular}
\caption{Turbulence characteristics of the velocity and magnetic fields used to construct $\mhdebd$.
%All quantities are in MKS units. 
$E_u(\kappa)$ and $E_b(\kappa)$ designate the velocity and magnetic 
field spectra, e.g., $E_u=\frac32 u^2 = \int E_u(\kappa) d\kappa$. Angular brackets 
denote a spatial average.}
\label{tableparamflow}
\end{center}
\end{table}
To construct $\mhdebs$ we modify the phases of $\tilde \bfb({\bf k})$ and $\tilde \bfu({\bf k})$
according to,
%
%\begin{align}
%\begin{equation}
\begin{eqnarray}
\alpha_i ({\bf k}) \rightarrow \alpha_i ({\bf k}) + \phi({\bf k}), \quad
\beta_i ({\bf k}) \rightarrow \beta_i ({\bf k}) + \varphi({\bf k}),
\label{randphase1}
\end{eqnarray}
%\end{equation}
%\end{align}
%
where $\phi({\bf k})$ and $\varphi({\bf k})$ are random numbers distributed uniformly
in the interval $[-\pi,\pi]$ and which satisfy $\phi({\bf k}) = - \phi(-{\bf k})$ and 
$\varphi({\bf k}) = - \varphi(-{\bf k})$. The electric field is then computed from Eq. (\ref{je}).
The above transformation ensures that 1) the phases of the new fields $\mhdebs$ 
contain a highly random part; 2) the spectra of ${\bf u}$ and ${\bf b}$ remain unchanged;
3) the divergence of $\bfb$ remains equal to zero; 4) the ideal MHD condition $\bfe \cdot \bfb = 0$
is preserved if $\eta = 0$, so that the randomization does not introduce parallel electric fields.

\subsection{Intrinsic Properties of the Accelerators $\mhdebd$ and $\mhdebs$}

The whole purpose of building an accelerator using DNS of the MHD equations
is to obtain an electromagnetic configuration that has more characteristic
features of turbulence than only the correct energy spectra. In this
section we describe some of the differences between $\mhdebd$ and $\mhdebs$
that are relevant from the point of view of turbulence and particle acceleration.
See \cite{verma04} for a recent review on MHD turbulence.

\begin{figure}[h!]
\centerline{
\plotone{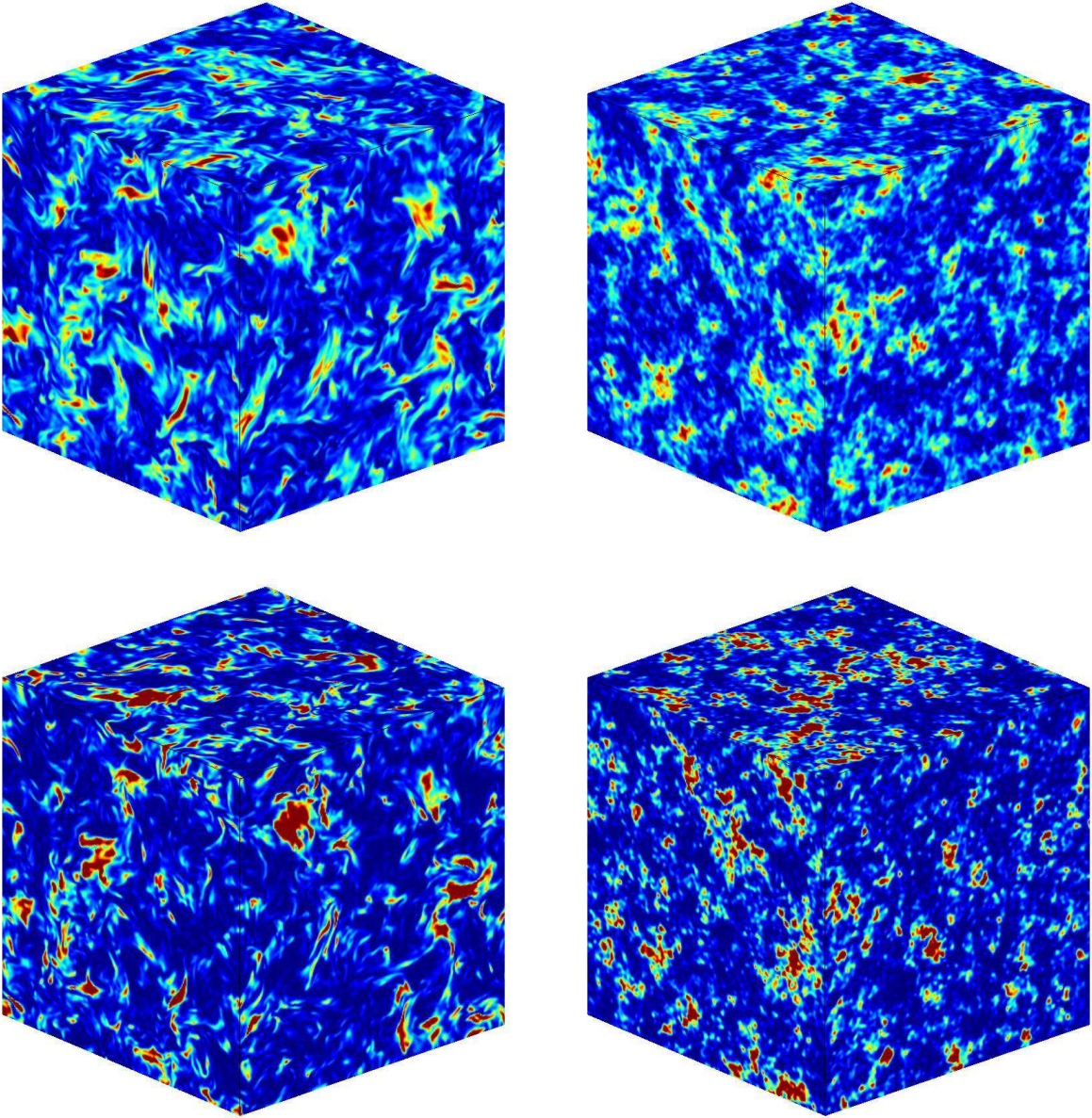}}
%\includegraphics*[width=210pt]{f1a.eps}			% NB: THE ASTERISK IS ESSENTIAL!
%\includegraphics*[width=210pt]{f1b.eps}}
% this was B_small.eps and Bscramble_small.eps
%\centerline{
%\includegraphics*[width=210pt]{f1c.eps}
%\includegraphics*[width=210pt]{f1d.eps}}
% this was E_small.eps and Escramble_small.eps
\caption{Magnetic (top line) and electric (bottom line) energy densities of the
$\mhdebd$ (left column) and $\mhdebs$ (right column) fields, similar as in
\cite{dmitruk03}. Blue and red regions indicate respectively low and high values.}
\label{becube_fig}
\end{figure}

Figure \ref{becube_fig} presents 3D views of the magnetic ($|\bfb|^2$) and electric 
($|\bfe|^2$) energy densities. Although this corresponds only to a
visual impression, it appears clear from the graphs that the
electromagnetic field composing $\mhdebd$ hosts more coherent, filamentary
structures than the one composing $\mhdebs$. In the latter, structures
appear noisier and span regions of smaller extent.
\clearpage
\clearpage
Also, it appears from Fig. \ref{becube_fig} that the electric field has more
extreme values (dark red) than the magnetic field, as already noticed by \cite{dmitruk03}.
More quantitative information on this can be 
obtained from the distributions (one-point functions) of individual field components. 
In the case of $\mhdebs$ it is observed, as expected, that $u_1$ and $b_1$ have a nearly 
Gaussian distribution (other components behave similarly). The distribution of $e_1$ 
(again for $\mhdebs$) is very close to a zero-th
order modified Bessel function of the second kind, which decays
exponentially for large values of its argument. This is again expected, since
at high Reynolds number $e_1$ is dominated by the term $- \bfu \times \bfb$
which behaves as the product of two Gaussian variates for $\mhdebs$.
In the case of $\mhdebd$, the distribution of the field components are similar
to their random-phase equivalents. Only for large values do they exhibit
moderately pronounced tails resulting from higher intermittency.
\clearpage
\begin{figure}[h!]
\epsscale{1}
\plottwo{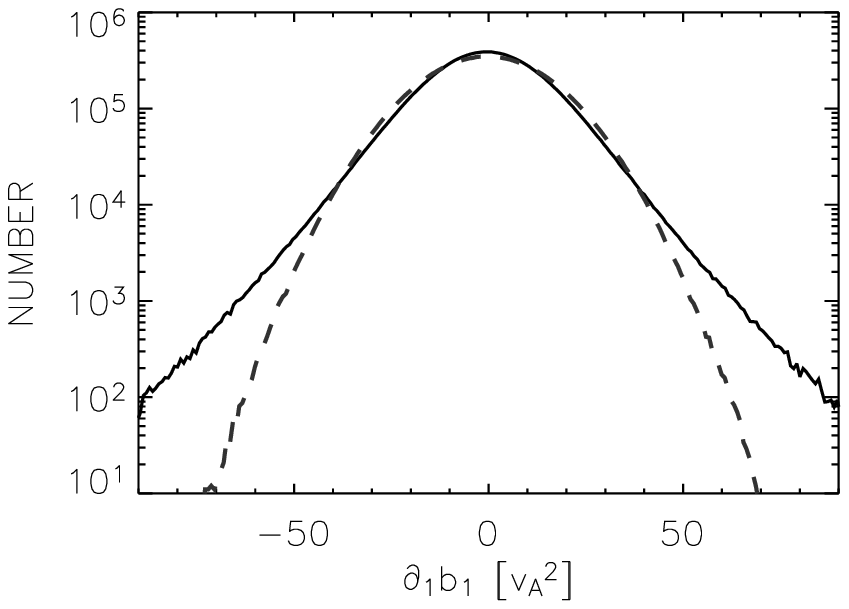}{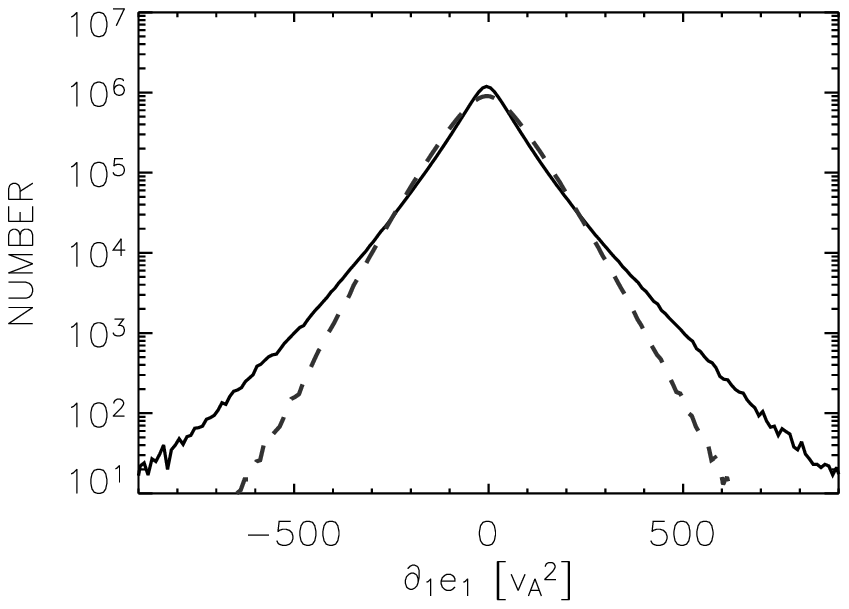}
% this was \plottwo{histodB_kaspar.eps}{histodE_kaspar.eps}
\caption{Histograms for the longitudinal 
derivatives of magnetic (left) and electric (right) field components. 
Solid -- DNS; dashed -- random phase.}
\label{histodBE_fig}
% created by pc4522:/scratch/bernard/histo_kasparEB.pro
\end{figure}
\clearpage
A better measure of coherency is contained in the distributions of the longitudinal derivatives
such as $\partial_1 b_1$ and $\partial_1 e_1$ (cf. \citealt{gotoh02}) (Fig. \ref{histodBE_fig}).
%
%, and some of their statistics are contained in table \ref{tableDistr}. 
%
Again, the 2nd and 3rd components behave similarly and are therefore not
shown. In the DNS case, both $\partial_1 b_1$ and $\partial_1 e_1$ depart
significantly from their random-phase equivalents. The distributions
exhibit a large number of extreme values (tails in the distribution). These
events allow coherent structures to be present in the signal since they
correspond to regions of space where sharp transitions  ``shaping" the signal 
can occur. The strong differences in this diagnostic between the
DNS and random phases case reflect the marked visual differences observed
in Fig. \ref{becube_fig}.
\clearpage
\begin{figure}[h!]
\epsscale{0.55}
\plotone{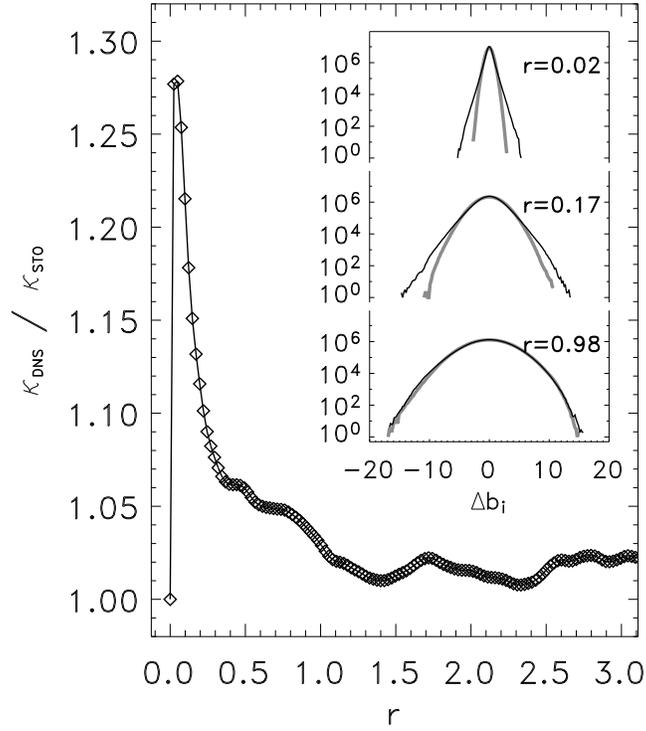}
% this was kurtosis_b.eps
\caption{Diamonds: the excess of the kurtosis of $\Delta b_i = b_i(\bfx+r{\bf \hat{e}}_i)-b_i(\bfx)$ 
between DNS and STO fields, averaged over $i$ = 1,2,3. Inlets: some representative histograms of 
$\Delta b_i$ (average over $i$=1,2,3); black and gray lines represent DNS and STO cases. See text.}
\label{kurtosis_fig}
% created by pc4522:/scratch/bernard/struc_funct_kaspar.pro
\end{figure}

\begin{figure}[h!]
\epsscale{0.5}
\plotone{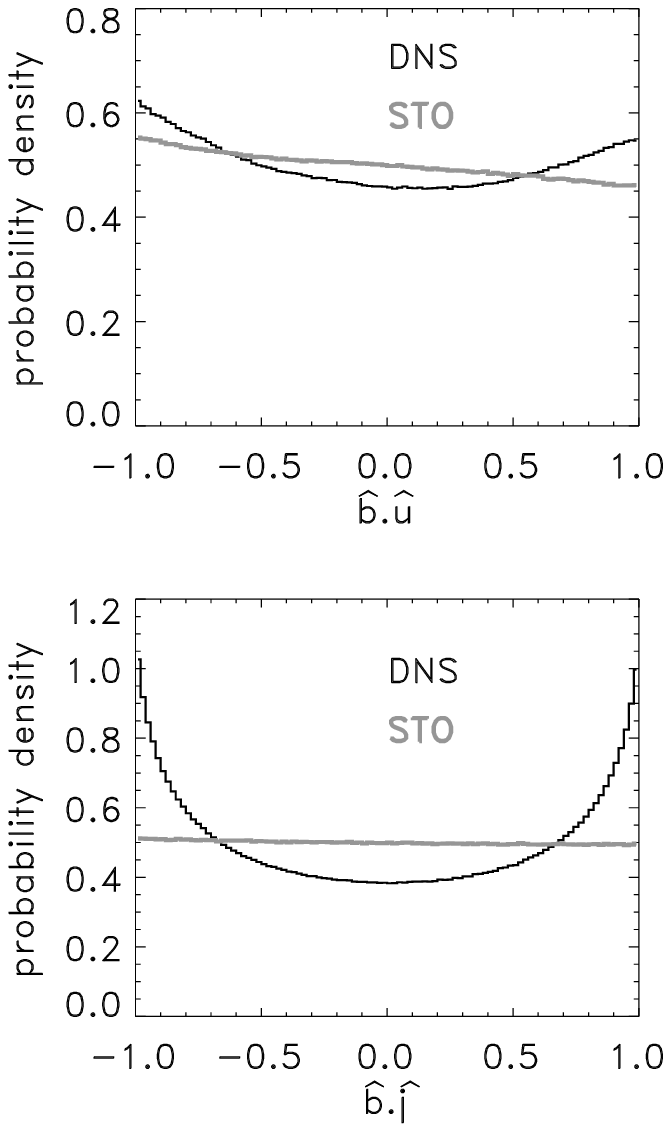}
% this was buj.eps
\caption{Distributions of the angle cosines between $\bfb$ and $\bfu$ (top), 
and $\bfb$ and $\bfj$ (bottom).}
\label{buj_fig}
% created by pc4522:/scratch/bernard/buj_kaspar.pro
\end{figure}
\clearpage
The coherent structures also manifest in the longitudinal structure functions
\citep{kolmogoroff41,mccomb90,frisch95,sorriso-valvo00,falkovich01,gotoh02,cho02} which are a natural
extension of the longitudinal derivatives discussed above. The magnetic longitudinal
structure functions are defined by $D_p(\bfr) = \langle | [\bfb(\bfr+{\bf
x}) - \bfb({\bf x})] \cdot {\bf \hat{r}}|^p \rangle$ and measure the
variation of the $p$-th moment of the longitudinal magnetic field increment
with distance $r$ (hat denotes a unit vector). Figure \ref{kurtosis_fig} (inlet) shows
the distribution of these increments along ${\bf \hat{r}}_i$, averaged over $i$ = 1,2,3.
With increasing $r$, the increments become decorrelated and their distribution becomes 
Gaussian. The transition can be monitored by means of the kurtosis $\kappa(r) = D_4(r)/D_2(r)^2$; 
Fig. \ref{kurtosis_fig} (diamonds) shows the DNS-to-STO kurtosis ratio as a
function of $r$. There is a clear systematic excess of the DNS kurtosis at small $r$,
and one may infer from Fig. \ref{kurtosis_fig} that on scales $r > 1$, the DNS and STO
fields behave similarly. In particular, the DNS field becomes independent at separations
which are long compared to the integral length-scale of $\bfu$ and $\bfb$ (Tab. \ref{tableparamflow}).

Another distinction between $\mhdebd$ and $\mhdebs$, which will prove helpful in understanding the
particle acceleration, is the relative orientation of $\bfb$ with respect to $\bfu$ and $\bfj$.
The nonlinear MHD equations (\ref{NSb}) and (\ref{NSu}) tend to align
$\bfb$ with $\bfj$ in order to minimize the $\bfj \times \bfb$ force. The (approximate, $Re_b$ = 283)
frozen-in condition tends, then, to align $\bfb$ with $\bfu$ \citep{biskamp93,dar98,debliquy05}. This preference 
for alignment is demonstrated in Fig. \ref{buj_fig}, showing the distribution of the orientation cosines 
$\bhat \cdot \uhat$ and $\bhat \cdot \jhat$ within the simulation cube. The DNS distributions (black line)
accumulate at the extreme values $\pm 1$, corresponding to $\pm 180^o$, more pronounced for 
$\bhat \cdot \jhat$ than for $\bhat \cdot \uhat$. After the phase randomization is applied (gray line),
this accumulation vanishes and the distribution of $\bhat \cdot \jhat$ (botom panel) become isotropic.
The residual anti-correlation between $\bhat$ and $\uhat$ (top panel, gray line), corresponding to
a negative cross-helicity, is dominated by few small-$|{\bf k}|$ modes, and is within statistical errors.
Note that for the mixed-field correlations, the coherent structures already manifest at the level 
of $\bfb \cdot \bfu$ and $\bfb \cdot \bfj$, whereas one has to go to higher order moments 
if $\bfb$ and $\bfu$ are separately considered (Fig. \ref{kurtosis_fig}).

\section{\label{orbit_sect}PARTICLE ORBITS}

\subsection{Physical Environment}

An ensemble of collisionless charged particles is amenable to the MHD description if 
their orbits can be described by the drift velocity $\bfe \times \bfb / |\bfb|^2 + {\bf v}_\parallel$, 
where `$\parallel$' refers to the local magnetic field. The drift velocity then equals the MHD bulk 
velocity $\bfu$. This description requires that the Larmor radius is much smaller than the
radius of curvature of the magnetic field, and that $v_\parallel$ is smaller
than the Alfv\'en velocity (e.g., \citet{littlejohn82,buechner89}). In the quiet 
solar corona, the ordering $v < v_A$ holds true for both electrons and ions. During 
solar flares, when turbulence arises, a small fraction of the particles
reaches relativistic energies, and it is this high-energy population which
is treated here as test particles. The non-energetic bulk provides the MHD
acceleration environment. The acceleration to relativistic energies is known
to be rather rapid. In impulsive flares, hard X ray signatures
of relativistic electrons appear within fractions of a second, and Gamma
ray signatures of relativistic protons -- if any -- appear within a few seconds
\citep{miller97}. The MHD structures, in contrast, evolve on a much longer
time scale.

The solar corona, in which acceleration is thought to occur, is a relatively
collisionless environment. Assuming a magnetic field strength of $10^{-2}$T, the
electron and proton gyro times are in the order of nanoseconds and microseconds,
respectively. In contrast, the ion-electron collision time is in the order of
tens of milliseconds if we adopt a typical temperature $T_e = T_i = 10^6$K
and density $10^{16}$ m$^{-3}$. The particles of the background plasma thus 
perform many gyrations before a Coulomb collision is encountered, and this
becomes even more true for high-energy test particles, because the collision
rate rapidly decreases with energy (\citealt{dreicer59, helander02}). While
the background plasma relaxes towards a Maxwellian on time scales of tens 
of milliseconds, a relativistic test proton remains virtually collisionless.

For particles which move much faster than the background plasma, the MHD
fields may be regarded as static. This requires $v \gg v_A$. In the solar
corona, $v_A$ is in the order of 1000 km/s, and the condition $v \gg v_A$ is fulfilled for
relativistic particles. However, acceleration begins at Alfv\'enic velocities, and the question 
rises whether a static approximation applies from the very beginning. This depends on the time
spent at low velocities. As outlined above, this time is observationally known to
be short compared to the MHD time scale in the real solar corona, and in our orbit 
simulations we will find that particles reach relativistic velocities within some $0.01
t_A$. The accelerated particles therefore spend most of the simulation in a state with 
$v \gg v_A$, and we set
\begin{equation}
\bfb({\bf x},t)=\bfb({\bf x},t_0),\
\bfe({\bf x},t)=\bfe({\bf x},t_0),\ \forall t
%\partial_t \bfe = \partial_t \bfb = 0
\label{snapshot}
\end{equation}
throughout the whole orbit simulation. The stationarity assumption (\ref{snapshot}) admits conservation 
laws which facilitate the numerical benchmarking and the physical understanding of the acceleration process.
In this sense, Eq. (\ref{snapshot}) also represents a simplifying working hypothesis.
%
%The stationarity assumption
%(\ref{snapshot}) is further supported since the particles gain energy mostly inside 
%the dissipation regions where $\eta \bfj$ is not negligible (cf \citealt{arzner05}).
%Therefore, the condition (\ref{snapshot}) must effectively only hold during visits to
%dissipation regions.

The test particles orbits are measured in the same units as the MHD fields.
The dimensionless equations of motion read
\begin{eqnarray}
\frac{d {\bf x}}{dt} 		& = & {\bf v} \label{dxdt} \\
\frac{d \gamma {\bf v}}{dt} 	& = & \alpha (\bfe + \bfv \times \bfb)
\label{dpdt} \, ,
\end{eqnarray}
where $\gamma = (1-v^2/c^2)^{-1/2}$ is the Lorentz factor, and the
effective charge $\alpha = \Omega t_A$ defines the particle time scales
relative to the MHD time scale \citep{dmitruk03}, with $\Omega$ the proper rms gyro
frequency. Assuming a coronal magnetic field of $10^{-2}$T and a density of
$10^{16}$m$^{-3}$, we have $v_A$ = 2182 km/s = 0.0073c (thus $c$=137.61 
in numerical units), and focusing on protons as test particles, we set $\alpha = 10^3$. 
For display purposes, time is measured in units of the (dimensionless) proper gyro time 
$t_C = 2 \pi / (\alpha \sqrt{\langle |\bfb|^2\rangle} = 3.6 / (\alpha v_A)$, corresponding
to about $5 \cdot 10^{-6}$ s in real time.
All test particles start out at random position and in random direction with initial 
velocity $v_0 = 0.2 v_A$, representative of protons with $T = 10^7$K. This ensures that 
the test protons can be considered as approximately collisionless, and that 
the initial Larmor radius is smaller than the numerical cell. The orbits are
exactly integrated, without recourse to gyrokinetic approximations. This
approach benefits from simplicity and rigor, but allows only relatively small
populations to be simulated over relatively short times. See Section \ref{summary_sect}
for further discussion of the limitations of the method.

\subsection{Numerical Implementation and Benchmarking}

Equations (\ref{dxdt}) and (\ref{dpdt}) are integrated by a 4th order
Runge-Kutta scheme with constant time step. The electromagnetic fields are
taken from a snapshot of the pseudo-spectral MHD simulation, transformed
to real space, and linearly interpolated to the actual particle positions. A
parallel (distributed memory) implementation is used to enhance the
computational power. The diagnostics includes energy histograms and the
second moments of momentum and space displacements.
\clearpage
\begin{figure}[h]
\epsscale{0.5}
\plotone{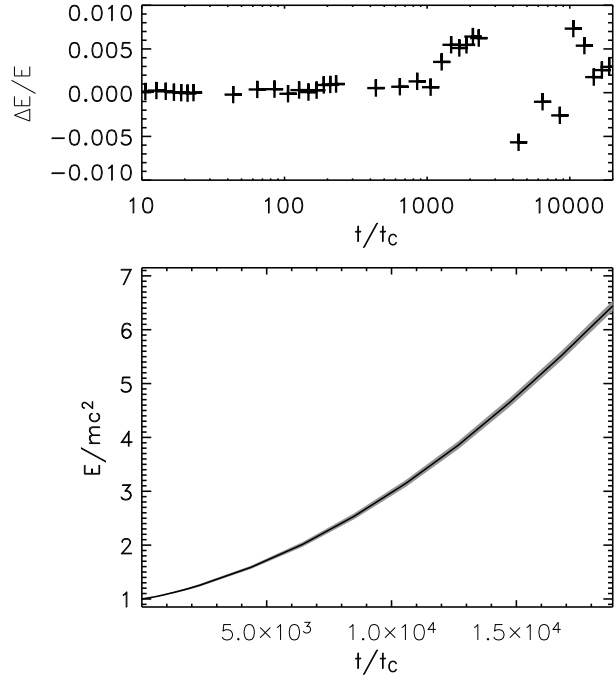}
% this was hamilton_kaspar.eps 
\caption{Benchmarking of energy conservation (Eq. \protect\ref{dH}). Top: relative error
along an example orbit. Bottom: ensemble average, with the shaded region representing the $\pm 5 \sigma$ 
deviations. Time is measured in units of the proper rms gyro time $t_C$.}
\label{num_conserve_fig}
% created by pc4522:/scratch/bernard/hamilton_kaspar.pro
\end{figure}
\clearpage
The static assumption (Eq. \ref{snapshot}) implies that the vector
potential evolves linearly with time, and that the scalar potential $\phi$
is independent of time. Therefore the identity
\begin{equation}
\gamma c^2 + \alpha \phi  = H_0 - \alpha \int dt \, \bfv \cdot \bfe_{\rm sol}
\label{dH}
\end{equation}
must hold, where $\bfe_{\rm sol}$ is the solenoidal ($\nabla \cdot \bfe_{\rm sol}$ 
= 0) part of the electric field and $H_0$ is the initial energy. Numerically,
$\phi$ and $\bfe_{\rm sol}$ are computed from the Fourier amplitudes
$\tilde{\bfe}({\bf k})$ by decomposition into parallel and perpendicular
components with respect to ${\bf k}$. To make the decomposition unique,
the potential is gauged such that $\langle \phi \rangle = 0$. 
Equation (\ref{dH}) can be used to assess the numerical accuracy and to choose an 
appropriate time step. To this end, the right hand side of Equation (\ref{dH}) is 
numerically integrated along the trajectories and compared to the lefthand side (Fig. 
\ref{num_conserve_fig}). The observed discrepancy $\Delta E$ depends on the time step 
$\Delta t$ and on the smoothness of the MHD fields; both small $\Delta t$ and smooth fields help
to fulfill Equation (\ref{dH}). In view of the fact that real acceleration
increases the particles' kinetic energy by orders of magnitudes, an rms
discrepancy of 1\% on the 2$\sigma$ level between the left- and righthand sides of Equation 
(\ref{dH}) at the end of the simulation is considered as acceptable. This is
achieved if $\Delta t < 0.01 \times (\alpha v_A)^{-1}$.

The Runge-Kutta integration scheme was also tested against a simpler
leapfrog scheme and it was shown that it provided a significant increase
in precision.
\clearpage
\begin{figure}[ht]
%\vspace{5cm}
\epsscale{1}
\plottwo{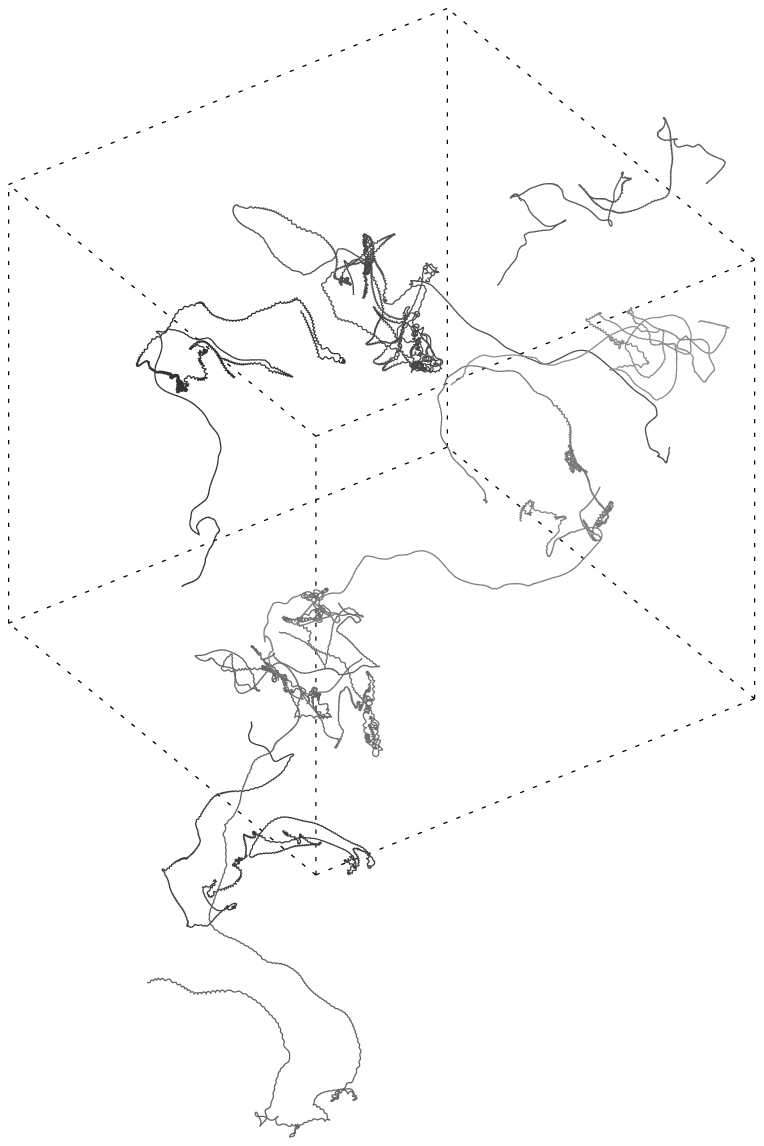}{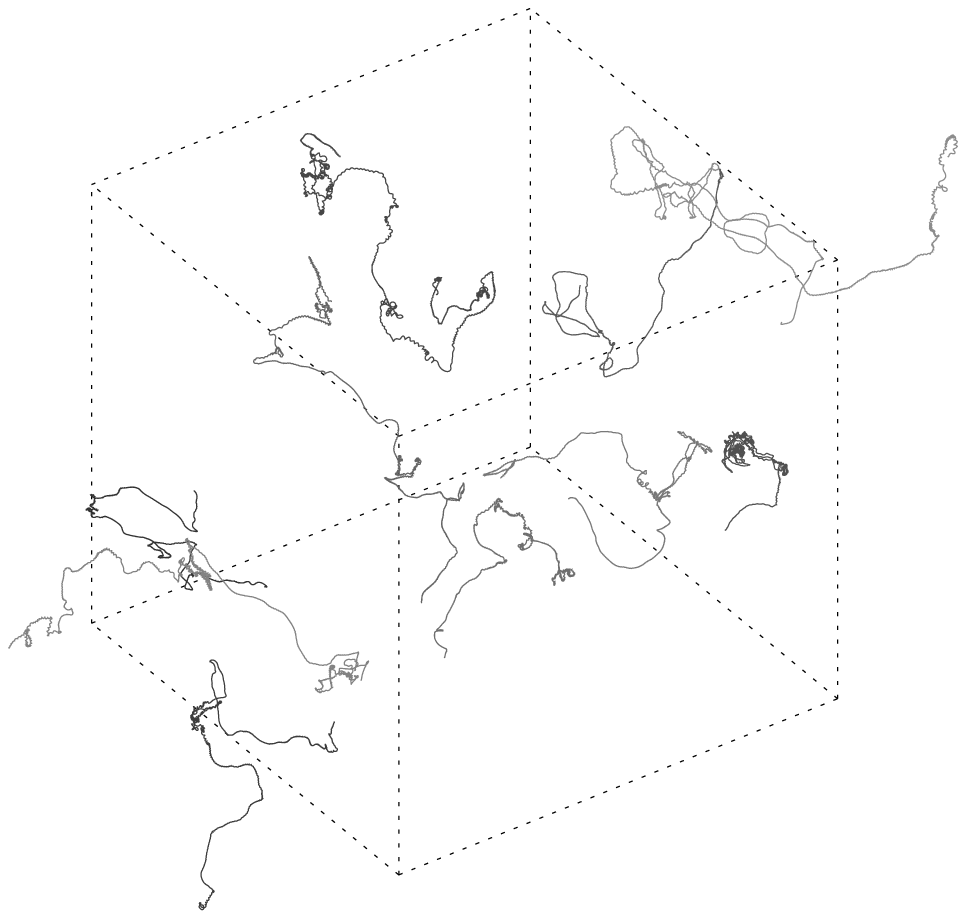}
% this was traj3D_10part_CORR.eps and traj3D_10part_STOC.eps
\caption{Real-space orbits in $\mhdebd$ (left) and $\mhdebs$ (right) fields
for $0 \le t/t_C \le 1.5 \cdot 10^3$. The gyration is not resolved in this view.}
%create dby pc4522:/scratch/bernard/traj3D_10part_kaspar.pro
\label{traj3d}
\end{figure}
\clearpage

\section{\label{result_sect}RESULTS}

In this Section we summarize the outcome of the particle simulations, focusing on
the second moments and correlation functions of the dynamical quantities ($\bfx, \bfp$).
The average over particles will be denoted by $\llangle ... \rrangle$ in order
to distinguish it from the spatial average $\langle ... \rangle$. The particle ensembles comprise 50.000 particles.
\clearpage
\begin{figure}[ht]
\epsscale{0.7}
\plotone{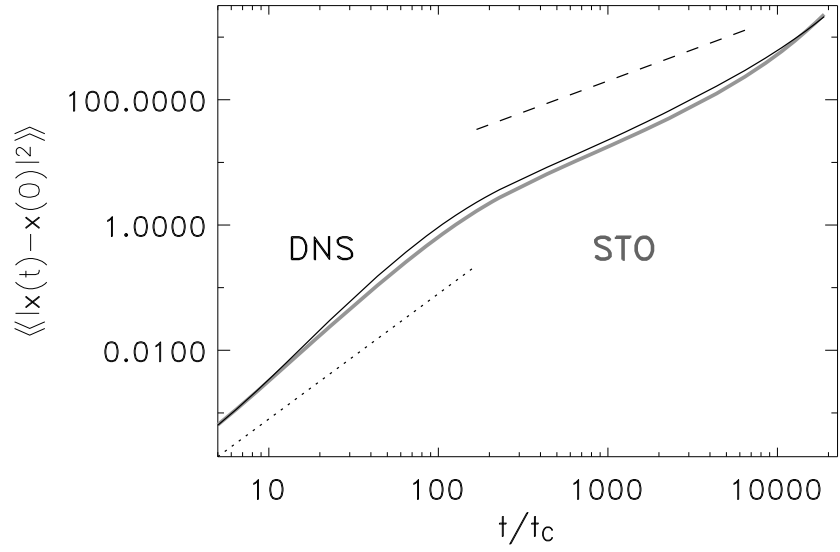}
% this was delta_x_kaspar.eps
\caption{Mean square displacement in correlated (black) and decorrelated (gray) fields.
The dashed and dotted lines indicate linear and quadratic growth.}
\label{delta_x_fig}
% created by pc4522:/scratch/bernard/delta_x_kaspar.pro
\end{figure}
\clearpage
\subsection{\label{x_diff_sect}Real-Space Diffusion}

A first qualitative difference between the $\mhdebd$ and $\mhdebs$ cases
concerns the intermittency of their real-space
trajectories. Figure \ref{traj3d} shows sample orbits in $\mhdebd$ (left)
and $\mhdebs$ (right) fields for times up to $t/t_C \le 1.5 \cdot 10^3$.
As can be noticed, the $\mhdebd$ orbits tend to have long sojourns between 
clusters where particles are temporarily trapped, and depart more from their
origin than the $\mhdebs$ trajectories. This tendency, observed for the few 
samples in Figure \ref{traj3d}, reflects in a full
population average of the mean square displacement (Fig. \ref{delta_x_fig}).
Here we see that the $\mhdebd$ average grows somewhat faster than the 
$\mhdebs$ average during the first 200 gyro times, from where on a
relative excess of about 30\% is preserved up to $t/t_C \sim 10^4$.
During this time interval, the mean-square displacement grows approximately
linearly with time (dashed). Later on, the
$\mhdebs$ curve overtakes the $\mhdebd$ curve, and grows again faster than
linearly with time. The bends observed in Fig. \ref{delta_x_fig} indicate
changes of acceleration regimes which will be discussed below.

The distance traveled by the test particles ultimately
exceeds the side length of the MHD simulation cube, and the
particles cross many periodic continuations of this cube. However, the
particle orbits themselves are not periodic, because opposite points like
$(x,y,0)$ and $(x,y,l_z)$ are not magnetically connected (this is a consequence of 
the fully 3-dimensional nature of the MHD simulation), and because the particles
may leave their magnetic field lines. Rather, the particle orbits have mixing
and ergodic properties. Therefore, the periodicity of the MHD fields is not
expected to affect the particle results.

\clearpage
\begin{figure}[ht]
\plotone{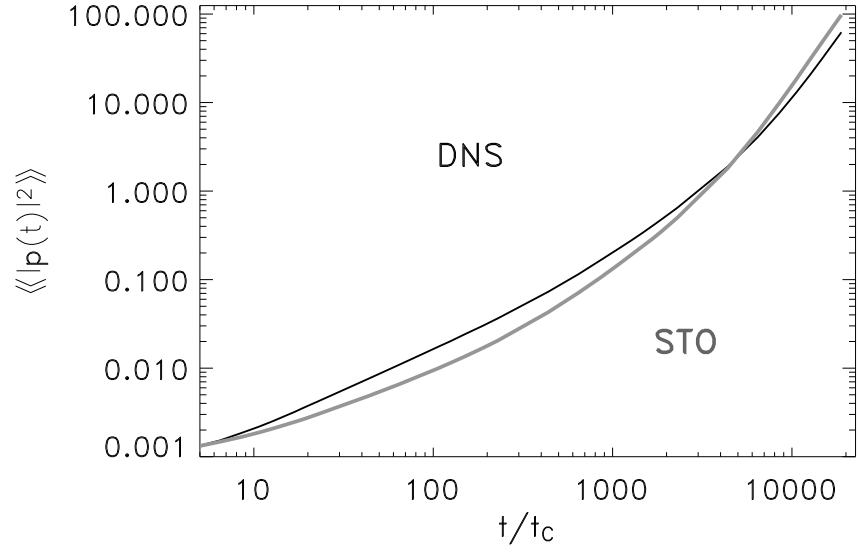}
% this was delta_p_kaspar.eps
\caption{Mean square momentum in correlated (black) and decorrelated (gray) fields.}
\label{delta_p_fig}
% created by pc4522:/scratch/bernard/delta_p_kaspar.pro
\end{figure}
\clearpage
\subsection{\label{p_diff_sect}Momentum Diffusion}

Next we consider the diffusion in momentum space (momentum ${\bf p} =
\gamma \bfv$ is a better indicator for diffusion than the velocity because
it is not limited by the speed of light). The evolution of the mean square momentum
is shown in Figure \ref{delta_p_fig} for both $\mhdebd$ and $\mhdebs$ ensembles. The
$\mhdebd$ curve grows faster at $10 < t/t_C < 10^2$, which is the typical particle 
crossing time of the coherent structures. Later, when many coherent structures have 
been visited ($t/t_C \sim 5 \cdot 10^3$), the $\mhdebd$ curve overtakes.
%The convex shape of the log-log curves in Fig. \ref{delta_p_fig} indicate that a scaling relation $p \sim t^\xi$ is not valid.
The average terminal ($t/t_C = 1.9 \cdot 10^4$) energy is $\llangle \gamma \rrangle = 6.4$
for the $DNS$ case and $\llangle \gamma \rrangle = 8.0$ for the STO case. Thus our
protons are accelerated to relativistic energies.
\clearpage
\begin{figure}[ht]
\plotone{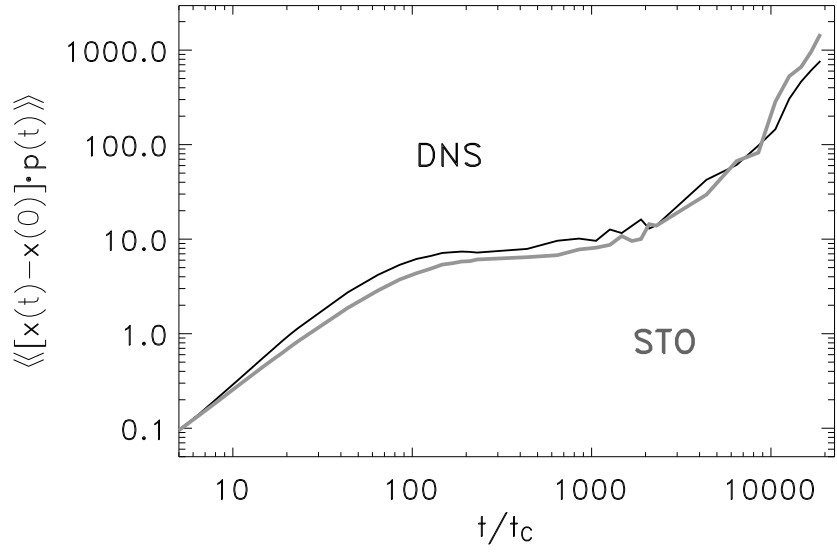}
% this was delta_xp_kaspar.eps
\caption{Displacement-momentum correlation.}
\label{delta_xp_fig}
% created by pc4522:/scratch/bernard/delta_xp_kaspar.pro
\end{figure}

\begin{figure}[ht]
\plotone{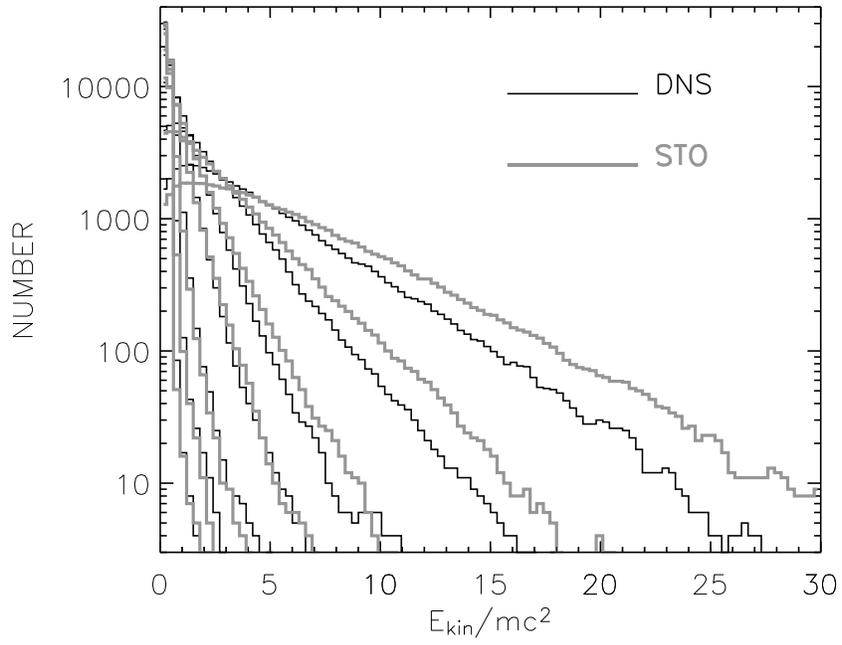}
% this washistoE.eps
\caption{Energy distributions at $t/t_C$ = 643, 1264, 2299, 4369, 6439, 10579, 16789 (left to right).}
\label{histoE_fig}
% created by pc4522:/scratch/bernard/histop_kaspar.pro
\end{figure}
\clearpage
\subsection{\label{xp_diff_sect}Momentum-Displacement Correlation}

The propagation of diffusion from momentum space (where $\mhdebd$ and $\mhdebs$ act) into real-space
is governed by the mixed correlation $\frac{d}{dt} \llangle |\bfx(t)-\bfx(0)|^2 \rrangle = 
2\llangle [\bfx(t)-\bfx(0)]\cdot \bfv(t) \rrangle$. This
quantity is always positive due to the statistical preference for particles having traveled 
away from the origin to have a velocity component in the same direction.
Since the coherent structures are able to channel the particles along filaments, they impose
a correlation between position and momentum, and we may expect $\llangle [\bfx(t)-\bfx(0)]\cdot \bfp(t) \rrangle$ 
to grow faster in $\mhdebd$ than in $\mhdebs$ during the first structure-travel time.
This is indeed the case, as shown 
in Fig. \ref{delta_xp_fig}. The plateau at intermediate times $100 \le t/t_C \la 2000$
indicates the regime of ordinary diffusion.
\clearpage
\begin{figure}[ht]
\plotone{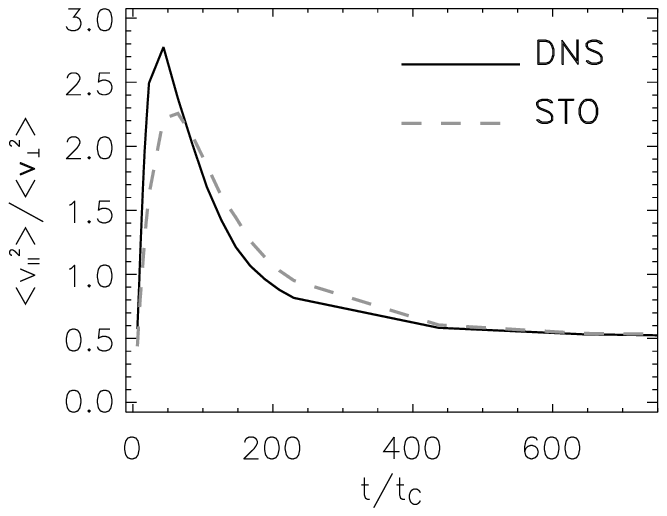}
% this was vparperp.eps
\caption{Particle anisotropy in the $\mhdebd$ and $\mhdebs$ fields.}
\label{vparperp_fig}
% created by pc4522:/scratch/bernard/delta_p_parperp_kaspar.pro
\end{figure}

\begin{figure}[ht]
\plotone{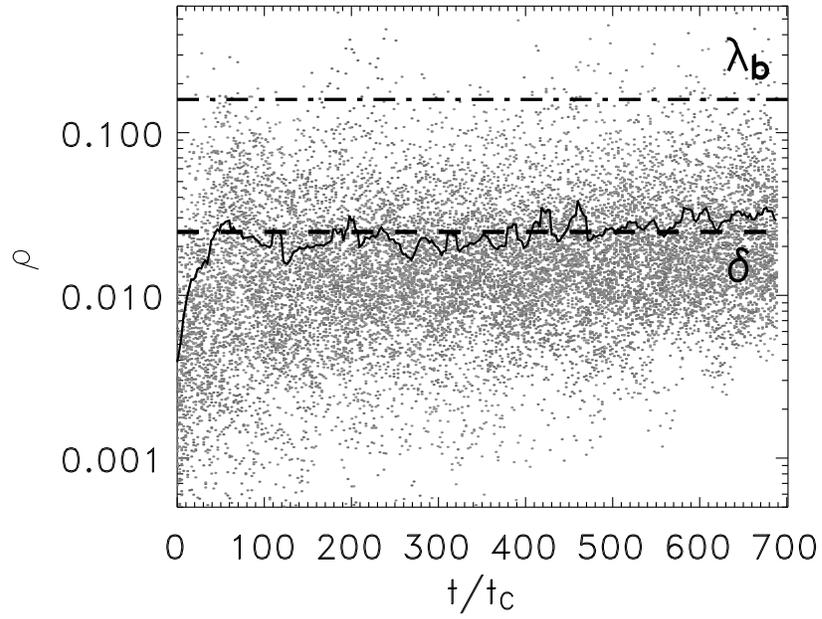}
% this was rho_linlog.eps
\caption{Orbital radius of curvature versus time. Dots: sample orbits; solid line: average.
For comparison, the numerical cell $\delta$ and the Taylor microscale $\lambda_b$ are also indicated.}
\label{rho_fig}
% created by pc4522:/scratch/bernard/rho_kaspar.pro
\end{figure}
\clearpage
\subsection{Energy Spectrum}

The evolution of the full energy distribution is shown in Fig. \ref{histoE_fig}. 
The particles start out at $t$=0 with kinetic energy 0.0037$mc^2$, and spread 
then to higher values. Note that for $t/t_C \la 3 \cdot 10^3$, 
the DNS fields are more efficient accelerators than the STO fields, 
while the converse holds true at later times. The spectral shape has slightly
positive (negative) curvature for $t/t_C \la 5 \cdot 10^3$ ($t/t_C \ga 5 \cdot 10^3$),
indicating that the distribution functions decay sub-exponentially (super-exponentially). 
When fitted by a power law $N(E) \sim E^{-\xi}$ in the sub-exponential range 
$400 < t/t_C < 2000$, power law indices $\xi = 4.5 ... 3.7$ are found.
Similar exponents have been observed in medium-size solar flares,
although for electrons, and smaller energies $E/m_ec^2 \la$ 0.2
\citep{grigis04}.

\section{\label{discussion_sect}ACCELERATION MECHANISM}

The ultimate cause of particle acceleration is the electric field which is now examined in some detail,
together with the microscopic nature of the particle orbits.
%in order to interpret the results of Sections \ref{x_diff_sect} and \ref{p_diff_sect}.
From a particle dynamics point of view, the main classification of $\bfe$ is into potential-solenoidal 
and parallel-perpendicular (to $\bfb$) components. From an MHD point of view, one may further 
distinguish between convective and dissipative fields.

\subsection{\label{solenoidal_sect}Solenoidal and Potential Electric Fields}

By virtue of the `snapshot' condition (Eq. \ref{snapshot}), the potential 
part $-\nabla \phi$ of the electric field does not contribute to acceleration on time scales which are long 
compared to the crossing time of local potential minima. However, the initially spatially uniform distribution 
of particles may become non-uniform at later times, resulting in a shift of the average kinetic energy which
would manifest in Figs. \ref{delta_p_fig} and \ref{histoE_fig}. This shift
is bounded by the magnitude of the potential fluctuations $\Delta \gamma = \alpha \Delta \phi/c^2 \la 0.08$,
which is small compared to the energies reached in the tail of the energy distribution (Fig. \ref{histoE_fig}). 
It is thus the solenoidal electric field $\bfe_{\rm sol}$ which is responsible for acceleration. 
From an MHD point of view, $\bfe_{\rm sol}$ is made up of the dissipative field $\eta \bfj$ plus parts of the convective field 
$- \bfu \times \bfb$. Numerically, the individual (rms) contributions are $e_{\rm sol, conv}^{\rm DNS}$ 
= 8.0, $e_{\rm sol, conv}^{\rm STO}$ = 9.35, and $e_{\rm sol, diss}^{\rm DNS}$ = $e_{\rm sol, diss}^{\rm STO}$ 
= 0.35. Thus the solenoidal electric field 
%(and in fact the whole $\bfe + \bfv \times \bfb$ force, see Fig. \ref{C_fig} middle) 
is somewhat weaker in the DNS than in the STO case, in contrast to the enhanced DNS acceleration efficiency
at $t/t_C \la 5 \cdot 10^3$.

\subsection{\label{parperp_sect}Parallel and Perpendicular Electric Fields}

The picture, though, changes if we distinguish between parallel and perpendicular components where
$e_{\rm sol, \parallel}^{\rm DNS} = 0.24 > e_{\rm sol, \parallel}^{\rm STO} = 0.20$.
Initially ($t$ = 0), the Larmor radius is smaller than the size of the magnetic inhomogeneities, 
so that the particles are in a gyrokinetic regime. The ratio of the (rms) forces on the particles is 
$|\bfv \times \bfb| \div |\bfe_\perp| \div |\bfe_\parallel| \sim 1.7 \div 12 \div 0.3$, so that the particles, 
typically, exhibit a strong $\bfe \times \bfb$ drift. The drift velocity $\bfv_d = \bfe \times \bfb / |\bfb|^2$ 
thus approaches $v_A$, and the maximal drift Larmor radius $r_L^* \doteq \frac{1}{2} \alpha |\bfe_\perp| (t_C/2)^2$ 
exceeds its initial (thermal) value $v_\perp/(\alpha b)$ by up to an order of magnitude.
% see pc4522:/scratch/bernard/rho_kaspar.pro (end of code)
This is a consequence of the strong turbulence assumption, where $v_A$ is given by the
perturbations and not by a (dominant) background magnetic fields.
Although the particles gain substantial perpendicular momentum during 
the first half gyration, this is approximately restored in the second half, and so forth, so that 
there is little net gain from $\bfe_\perp$. The parallel field $\bfe_\parallel$, however, yields direct 
parallel acceleration. As a consequence, the velocity distribution becomes anisotropic (Fig. \ref{vparperp_fig}), 
with a maximum anisotropy reached at $t/t_C \sim 50$. At later times, the velocities isotropize. One reason for this
is the (reversible) conversion of parallel into perpendicular momentum at converging magnetic field lines.
This mechanism is supported by the observation that $\llangle |\bfb|^2 \rrangle$
is slightly ($0.04^2$) enhanced during the isotropization phase $30 \la t/t_C \la 300$,
as would be expected when particles penetrate high-$|\bfb|$ domains.
% see pc4522:/scratch/bernard/D_kaspar.pro
%Since converging field lines are encountered independently for different particles and thus cancels in the average. 
%The STO field, with its blurred inhomogeneities and softer $|\bfb|$ gradients, yields a somewhat 
%less efficient conversion than the DNS field.

A second cause of the isotropization observed in Fig. \ref{vparperp_fig} is the fact that the 
gyrokinetic approximation breaks down with increasing energy. In order to monitor the transition 
from gyrokinetic to eventually free orbits we compare the orbital radius of curvature $\rho = 1/K$
to the characteristic scales of the magnetic field, where
$K^2 = (|{\bf \dot{x}}|^2 |{\bf \ddot{x}}|^2 - |{\bf \dot{x} \cdot \ddot{x}}|^2)/|{\bf \dot{x}}|^3$.
The evolution of $\rho$ is shown in Fig. \ref{rho_fig} for a sample of 50 particles (dots), with the
average $\llangle \rho \rrangle$  marked by solid line. Initially, $\llangle \rho \rrangle$ is smaller than the 
Taylor microscale $\lambda_b$ and also smaller than the numerical cell $\delta$ (the
electromagnetic fields are interpolated to the actual particle position). It then rapidly 
grows during $0 < t/t_C \la 50$, where it approaches the cell size and the growth slows down.
The quantity $r_L^*$ defined in the foregoing paragraph is found equal to the 
maximum of $\rho$ during the first gyration within $\pm$10\%. Based on Fig. \ref{rho_fig} and the criterion 
$\llangle \rho \rrangle > \delta$ one may infer that, on average, the gyrokinetic regime is left 
after a few hundred gyro times. At this point the character of the orbit and the acceleration mechanism
change.

A picture thus arises where acceleration is initiated by the parallel component of the
solenoidal electric field (`injection'), and is then taken over by the full non-potential electric field. 
While the parallel component is somewhat larger in $\mhdebd$, the total solenoidal field is larger
in $\mhdebs$. Thus the DNS fields are initially stronger accelerators, but the STO fields become
more efficient at a later (non-adiabatic) stage. In both cases, the parallel electric field 
is entirely dissipative. The initial (parallel) acceleration phase corresponds to the
first phase of Fig. \ref{delta_x_fig} ($t/t_C \la 200$).

The fact that the parallel field is weakened by the phase randomization,
while the perpendicular electric field is strengthened, is easily understood from Fig. \ref{buj_fig}.
The non-linear MHD alignment of $\bfb$ with $\bfj$ favours small $|j_\parallel|$; this preference is lost
in the phase randomization so that $|j_\parallel|$ increases. Similarly, the MHD alignment of
$\bfu$ and $\bfb$ is lost in the phase randomization, and $|\bfu \times \bfb|$ increases. Since 
the convective field numerically dominates the dissipative one,
%($Re \sim Re_b \sim 300$; see Sect. \ref{solenoidal_sect} for rms values of $\bfe$), 
the solenoidal electric field
as a whole increases when the phase randomization is applied to $\bfu$ and $\bfb$.

\clearpage
\begin{figure}[ht]
\epsscale{0.7}
\plotone{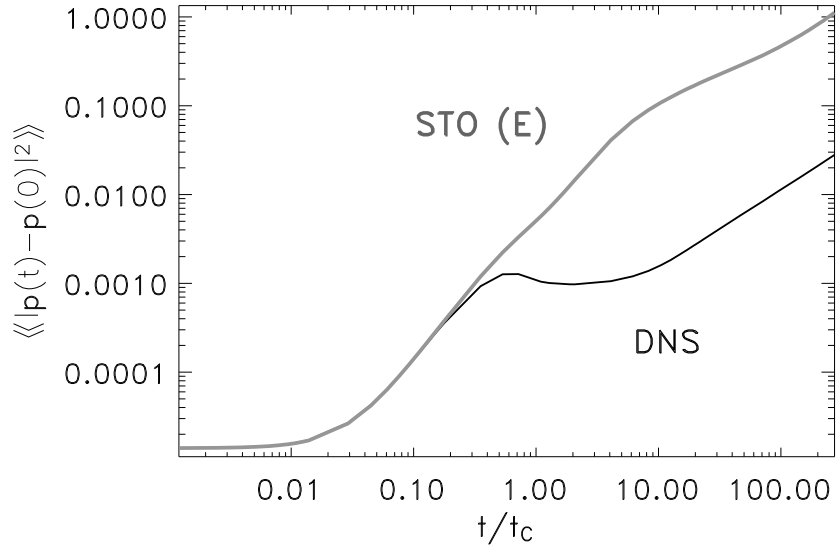}
% this was delta_p_Escramble.eps
\caption{Similar to Fig. \protect\ref{delta_p_fig}, but with $\bfe$ directly phase randomized (instead of 
computing it from the phase randomized $\bfb$ and $\bfu$). The time axis spans a smaller range than in Fig. 
\protect\ref{delta_p_fig}.}
\label{delta_p_Escramble_fig}
% created by pc4522:/scratch/bernard/delta_p_Escramble.pro
\end{figure}
\clearpage

\subsection{\label{direct_e_sect}Direct Phase Randomization of ${\bf E}$}

The crucial role of parallel electric fields can also be demonstrated by an alternative simulation setup, 
where $\bfe$ is directly phase-randomized rather than computed from the phase-randomized $\bfu$ and $\bfb$.
The ideal MHD condition $\bfe \cdot \bfb = 0$ is then broken even in the absence of
resistivity, while the power spectral densities of $\bfb$ and $\bfe$ remain unchanged. We may thus ask what
happens to the particles in such a directly phase-randomized electric field. This procedure 
has a certain interest on its own because of its implication on random-phase Fokker-Planck 
descriptions of stochastic acceleration.

It turns out that the direct phase randomization of $\bfe$ leads to a dramatic enhancement
of acceleration. Figure \ref{delta_p_Escramble_fig} shows the evolution of $\llangle |\bfp(t)-\bfp(0)|^2 \rrangle$
for the DNS and directly phase-randomized cases. The curve corresponding to direct phase
randomization of $\bfe$ is labeled STO (E). As can be seen, the DNS and 
STO (E) curves separate after about half a gyro period, and the direct phase
randomization of the electric field yields energies which are two orders of magnitude above 
the $\mhdebs$ results. This dramatic enhancement of acceleration is entirely due to (unphysical) parallel 
electric fields. The local maximum of the DNS curve at $t_C/2$ marks the first half gyro phase, where the
velocities are, on average, reversed. At this point, the particles in the directly phase-randomized electric
field have already gained super-Alfv\'enic velocities so that the effect of gyration is no
longer visible.

\section{\label{summary_sect}SUMMARY AND DISCUSSION}

We have performed a numerical study of test particle acceleration in strong
MHD turbulence ($\bfb, \bfu$) and its phase-randomized version, with the electric field given by $\bfe = 
-\bfu \times \bfb + \eta \bfj$. It is found that the dynamical alignment of $\bfb$ with $\bfj$ and $\bfu$ 
in true MHD affects the division of the electric field into parallel and perpendicular components, 
in the sense that parallel (resistive) components are favoured and perpendicular (convective) 
components are disfavoured. As a consequence, the true MHD turbulence provides more efficient direct 
acceleration of super-Dreicer particles than its phase-randomized version.
If the particles reach sufficiently high energies to leave the gyrokinetic regime,
so that they can become accelerated by the solenoidal part of the $\bfu \times \bfb$ force,
the dynamical alignment of $\bfu$ with $\bfb$ yields weaker acceleration than predicted
from the phase-randomized fields. For collisionless protons in MHD turbulence with magnetic 
Reynolds number $\sim$ 300, this is the case for protons after a few 
1000 gyro times. Numerically, the acceleration 
efficiencies of the true MHD and phase-randomized fields differ by factors of order two.
It is expected that the discrepancy would be larger for super-Dreicer electrons,
since these remain longer in the adiabatic regime where the enhancing effect of $(\bfb,\bfj)$ 
alignment is not yet competed with the diminishing effect of $(\bfb,\bfu)$ alignment. 

With regard to solar flares envisaged as a astrophysical application, 
a few words of caution are appropriate. The present simulation 
involves only dimensionless quantities such as $r_L/\lambda_b$ or $r_L / \delta$. In order to
fix the absolute physical scaling, and since we were interested in the possibility of non-adiabatic 
acceleration, we have chosen protons as test particle. Using typical values of the coronal
magnetic field strength, and taking the resolution of the simulation into account, the
protons become non-adiabatic ($r_L > \delta$) at $v \ga 0.6 \, c$, and reach energies as
high as 10 GeV (10 proton rest masses). Such high energies are commonly not observed; the highest
energies from in-situ observations in the interplanetary space are
about 100 MeV/nucleon \citep{reames92}, and proton energies inferred from pion decay
photons reach up to a few 100 MeV/nucleon and occasionally up to a few GeV \citep{kanbach93}.
(Only in the very largest flares such as of October 28, 2003, ground-based detections of relativistic 
protons and neutrons show energies up to 10 GeV \citep{bieber04}.) Also, the simulated
acceleration is very rapid; assuming $\langle|\bfb|^2\rangle \sim (100G)^2$, the simulation duration
is only 0.1s in real time. There are two main reasons for the violent acceleration in
the present simulation: the neglect of collisional and 
radiative losses, and the freedom in the absolute scaling of the simulation system. The
latter is chosen such that one numerical cell is about 200 initial ($|\bfv|=v_0$) 
Larmor radii; this corresponds to $\delta \sim$ 100m in real space. The resulting steep
gradients -- which are not detectable by present-day observations --, 
together with a larger-than-coronal resistivity ($Re_b \sim$ 300), yield 
highly efficient accelerators, which allow us to rigorously follow the 
particles to relativistic energies before becoming swamped by accumulated numerical 
errors (and running out of computing time). The simulation system is thus affected
by computational considerations, and presumably over-estimates the absolute coronal 
acceleration rate. However, the relative influence of dynamical MHD alignment on particle 
acceleration should be correctly reproduced.

Another caveat stems from the test particle nature of the present simulation, which does
not allow to address the question of absolute numbers of accelerated particles.
The test particles are considered as resulting from a runaway process which is out
of the scope of this simulation, so that we cannot directly relate their number 
to the number of thermal background particles. However, it is clear that the test particles must not
gain more energy than is available in the MHD fields. This sets an upper limit on their
number, and also on the duration of the simulation. (As the simulation does not 
include loss processes, the test particles can gain arbitrary large energy.) In our 
simulations, the test protons pick up Alfv\'enic velocities within one gyration. This brings them to
kinetic energies of the ambient (low-beta) coronal plasma, which is assumed to be in 
equipartition, $\langle |\bfu|^2 \rangle = \langle |\bfb|^2 \rangle$. In the course of 
the simulation, the kinetic energy of the test particles increases by a factor $10^4$.
This should not exceed a small fraction (say, $\epsilon = 10^{-2}$) of the magnetic field energy,
such as not to perturb the MHD configuration. Therefore the test particles represent a fraction
$f = 10^{-6}$ of all particles only. This fraction is small, but not unreasonably small for 
runaway protons, which must be substantially faster than the electrons in order for
runaway to occur \citep{dreicer60}. It is also in agreement withe the observation that
most solar flares do not show signature of high-energy protons (but only
of high-energy electrons). However, if the test particles are to represent more than 
$10^{-6}$ of all particles, then the acceleration should be stopped at earlier times.
Since it is found that the test particle energy increases roughly linearly with time,
the maximum simulation duration is $t_{\rm max}/t_C \sim \epsilon/f$.
In fact, it might happen that the MHD description brakes down after a few 100 gyro
times, so that the non-adiabatic regime where the $\mhdebs$ fields become 
more efficient accelerators is never reached.

Returning to the more technical aspects, and motivated by the findings of Sect. \ref{direct_e_sect}, 
we would like to stress that violation of the ideal MHD condition $\bfb \cdot \bfe = 0$
by random-phase approximations unavoidably leads to artificial particle acceleration.
If, in theory or numerics, a turbulence proxy is generated by superimposing Alfv\'en modes
$(\delta \bfb_\bfk,\delta \bfe_\bfk)e^{i\bfk . \bfx - i (\bfk .\bfb_0)t}$ with $\delta \bfe_k = 
\bfb_0 \times \delta \bfb_\bfk$, then a parallel electric field occurs which is of second order
in $\delta b$. For example, if the $\delta \bfb_\bfk$ are random isotropic in the 
plane perpendicular to $\bfb_0$, with orientation uncorrelated to $\bfk$, then $\langle 
(\bfe \cdot \bfb)^2 \rangle  = \frac{1}{4} b_0^2  \langle (\delta b)^2 \rangle^2$; for
$\delta b \ll b_0$, the parallel electric field has thus standard deviation 
$\frac{1}{2} \langle (\delta b)^2 \rangle$. Although this is small compared to the 
perpendicular electric field ($e_\parallel/e_\perp \sim \delta b / b_0$), it may 
give rise to secular growth of the parallel particle momentum. Therefore, 
stochastic acceleration in turbulence proxies constructed from superimposed 
Alfv\'en waves might in fact over-estimate the acceleration in true MHD turbulence. 

%On the other hand, parallel electric fields can also arise from compressional modes
%such as ion-acoustic waves (as observed by \cite{gurnett79} in the solar wind).

Finally, we point out that, although the coherent structures do not manifest in 
the two-point functions (or power spectral densities) of $\bfb$ and $\bfe$ alone, they contribute 
to their mixed correlations. This fact can, in principle, be taken into account in Langevin- or 
Fokker-Planck formulations of stochastic acceleration (e.g., \citealt{brissaud74}; \citealt{vankampen81}; 
\citealt{risken89}; \cite{urch77,urch91}; \citealt{gardiner97}; \cite{schlickeiser98,schlickeiser02}; 
\citealt{falkovich01}), which deal with the transfer of spatial correlators of the electromagnetic fields 
into the particle diffusion coefficients. Given the correlations $\langle b_i(0) \, b_j (\bfr) \rangle$, 
$\langle b_i(0) \, e_j (\bfr) \rangle$, and $\langle e_i(0) \, e_j (\bfr) \rangle$, one may compute, for example, 
the momentum diffusion coefficient $D_{pp} = \int_0^\infty \llangle \delta \bff(0) \cdot 
\delta \bff(\tau) \rrangle \, d \tau$ along the drifting gyrocentre $\bfx_0(\tau) = \bfx(0) + v_\parallel(0) \bhat \tau + \bfv_d \tau$ 
(Sect. \ref{parperp_sect}), where $\delta \bff = \bfe (\bfx) + \bfv \times \bfb(\bfx)$ for strong
turbulence. The quantity $D_{pp}$ may then be evaluated numerically, as follows. First we choose
a large ($10^7$) number of random initial data ($\bfx(0),\bfv(0)$) with uniform position $\bfx(0)$ and
isotropic velocity of fixed modulus $|\bfv(0)| = v_0 = 0.2 v_A$, like for the simulated test particles. 
Then, for each ($\bfx(0),\bfv(0)$) the instantaneous (straight)
gyrocentre orbit $\bfx_0(\tau)$ is computed, and the electromagnetic force 
$\delta \bff(\tau)$ is probed along this orbit. The average of $\llangle \delta \bff(0) \cdot 
\delta \bff(\tau) \rrangle$ over all particles (initial data) is performed, and the integral in the definition of $D_{pp}$ is
approximated by a sum of sufficiently resolved increments. Proceeding this way, we obtain  $D_{pp}$ = 0.03 for $\mhdebd$ and 
$D_{pp}$ = 0.018 for $\mhdebs$. 
% see pc4522:/scratch/bernard/D_kaspar.pro
The ratio of these values reflects the excess of $\llangle p^2 \rrangle_{\rm DNS}$
over $\llangle p^2 \rrangle_{\rm STO}$ observed in Fig. \ref{delta_p_fig} for 
$t/t_C \la 1000$. A detailed study of the modification of $D_{pp}$ by the cross-correlation
$\langle b_i(0) \, e_j (\bfr) \rangle$ has, to the authors knowledge, not yet been carried out.

\acknowledgments
The authors thank R. Kallenbach and M. Verma for helpful discussions.
L.V. is supported in part by the Research Training Network (RTN) 'Theory, Observation and
Simulation of Turbulence in Space Plasmas', funded by the European
Commission (contract No.HPRN-eT-2001-00310).

\end{document}